\documentclass[a4paper]{article}

\usepackage[utf8]{inputenc}
\usepackage{subfigure}
\usepackage{a4wide}

\usepackage{amssymb}
\usepackage{amsmath,amsthm}
\usepackage{mathtools}
\usepackage{enumerate}
\usepackage{tikz}
\usetikzlibrary{arrows,shapes,bending}
\usepackage{stmaryrd}
\usepackage{nicefrac}
\DeclareGraphicsExtensions{.png,.pdf,.jpg}

\usepackage{wrapfig}

\usepackage{algorithm}
\usepackage{etoolbox}\AtBeginEnvironment{algorithmic}{\footnotesize} 
\usepackage[noend]{algpseudocode}
\usepackage{mathrsfs}
\usepackage{caption}
\usepackage{url}

\newcommand{\A}{\mathcal{A}}

\newcommand{\IN}{\mathbb{N}}

\newcommand{\BigO}{\mathcal{O}}

\newcommand{\pari}{\texttt{par}}

\newtheorem{theorem}{Theorem}
\newtheorem{definition}[theorem]{Definition}
\newtheorem{proposition}[theorem]{Proposition}
\newtheorem{lemma}[theorem]{Lemma}

\title{Beyond series-parallel concurrent systems: the case of arch processes\thanks{
			This research was partially supported by the ANR MetACOnc project 
			ANR-15-CE40-0014.}}

\author{Olivier Bodini\thanks{Laboratoire d'Informatique de Paris-Nord,
			CNRS UMR 7030 - Institut Galil\'ee - Universit\'e Paris-Nord,
			99, avenue Jean-Baptiste Cl\'ement, 93430 Villetaneuse, France. 
			\texttt{Olivier.Bodini@lipn.univ-paris13.fr}}		,\ 
		Matthieu Dien\thanks{Institute of Statistical Science, Academia Sinica,
					Taipei 115,	Taiwan.
				\texttt{dien@stat.sinica.edu.tw}}, \
		Antoine Genitrini\thanks{Laboratoire d'Informatique de Paris 6,
			CNRS UMR 7606 and Sorbonne Univiersité,
			4 place Jussieu, 75005 Paris, France.
			\texttt{Antoine.Genitrini@lip6.fr}} \ and
		Alfredo Viola\thanks{Universidad de la Rep\'ublica, Uruguay.	
			\texttt{viola@fing.edu.uy}.}
		}

\date{\today} 

\begin{document}
	
\maketitle 

	\begin{abstract}
		In this paper we focus on concurrent processes built on \emph{synchronization}
		by means of \emph{futures}. This concept
		is an abstraction for processes based on a main execution thread but
		allowing to delay some computations.
		The structure of a general concurrent process with futures is
		more or less a directed acyclic graph. Since the quantitative study
		of such increasingly labeled graphs (directly related to processes) seems out of reach, we
		restrict ourselves to the study of \emph{arch processes}, a simplistic
		model of processes with futures. They are based on two
		parameters related to their sizes and their numbers of arches. 
		The increasingly labeled structures seems not to be specifiable in the
		sense of Analytic Combinatorics, but we manage to derive 
		a recurrence equation for the enumeration.\\
		For this model we first exhibit an \emph{exact} and an \emph{asymptotic} formula for the number of
		runs of a given process.
		The second main contribution is  composed of an \emph{uniform random sampler}
		algorithm and an \emph{unranking} one that allow efficient
		generation and exhaustive enumeration of the runs of a given arch process.
	\end{abstract}


	\section{Introduction}

Our study consists in the increasing labeling of combinatorial
structures, which tightly relates to the notion of behaviors of concurrent
processes. We conduct this study by using tools of
Analytic Combinatorics. This work is a part of a long time project to
understand the so-called \emph{combinatorial explosion}
phenomenon about the number of runs (or executions) of concurrent processes. In
previous works the authors studied tree-like processes \cite{BGP16}, tree-like processes with
non-deterministic choice \cite{BGP13} and Series-Parallel processes
\cite{diamonds16,BDGP17}.

The main common idea consists in modeling a concurrent
process as a partial order over the atomic actions of the process.
Thus some precedence relations describe the process.
In this way the runs of the process correspond to linear
extensions of the poset.
Then, we reinterpret this modelization in term of combinatorial structures
(trees, Series-Parallel graphs, directed acyclic graph, \dots)
where increasing labelings are in one-to-one
correspondence with the runs of the process.
Until the present work all the structures were
\emph{decomposable}, in the sense of \cite{FS09}, by recursive specifications.
For each of these families the objectives have always been the same:
understanding the growth of the number of runs for large
concurrent processes, which means understanding the combinatorial
explosion phenomenon;
and tuning efficient algorithm for the uniform random
generation of increasing labelings which is a practical way to
circumvent the combinatorial explosion phenomenon (see for example~\cite{GrSm04}).

\begin{wrapfigure}[18]{r}{4.7cm}
	\begin{center}
		\begin{tikzpicture}[xscale=0.5, yscale=0.5] 
  
  \node (a1) at (0,5) {$\prescript{a_1}{\phantom{f}}\bullet_{\;\;\;}$};
  \draw[->,>=latex,semithick] (a1) arc[radius=5, start angle=90, end angle=110] node (a2) {$\prescript{a_2}{}\bullet_{\;\;\;\;}$};
  \draw[->,>=latex,semithick] (a2) arc[radius=5, start angle=110, end angle=130] node (a3) {$\prescript{a_3}{}\bullet_{\;\;\;\;}$};
  \draw[dotted,->,>=latex,semithick] (a3) arc[radius=5, start angle=130, end angle=160] node (al) {$\prescript{a_k}{}\bullet_{\;\;\;\;}$};
  \draw[->,>=latex,semithick] (al) arc[radius=5, start angle=160, end angle=170] node (x1) {$\prescript{x_1}{}\bullet_{\;\;\;\;}$};
  \draw[dotted,->,>=latex,semithick] (x1) arc[radius=5, start angle=170, end angle=190] node (xr) {$\prescript{x_{n-k}}{}\bullet_{\;\;\;\;\;\;\;\;}$};
  \draw[->,>=latex,semithick] (xr) arc[radius=5, start angle=190, end angle=200] node (c1) {$\prescript{c_1 \;}{}\bullet_{\;\;\;\;}$};
  \draw[->,>=latex,semithick] (c1) arc[radius=5, start angle=200, end angle=220] node (c2) {$\prescript{c_2 \;\;}{}\bullet_{\;\;\;\;\;\;}$};
  \draw[->,>=latex,semithick] (c2) arc[radius=5, start angle=220, end angle=240] node (c3) {$\prescript{\phantom{f}}{c_3}\bullet_{\;\;}$};
  \draw[dotted,->,>=latex,semithick] (c3) arc[radius=5, start angle=240, end angle=270] node (cl) {$\prescript{\phantom{f}}{c_k}\bullet_{\;\;}$};
  
  \draw[<-,>=latex,semithick] (c1) arc[radius=9.7, start angle=-60, end angle=-10] node (a11) {};
  \draw (c1) arc[radius=9.7, start angle=-60, end angle=-25] node (b1) {$\prescript{b_1}{}\bullet_{\;\;\;\;}$};	
  \draw[-<,>=latex,semithick] (c1) arc[radius=9.7, start angle=-60, end angle=-23] node (bb1) {};
  \draw[<-,>=latex,semithick] (c2) arc[radius=9.7, start angle=-40, end angle=10] node (a22) {};
  \draw (c2) arc[radius=9.7, start angle=-40, end angle=-5] node (b2) {$\prescript{b_2}{}\bullet_{\;\;\;\;}$};
  \draw[-<,>=latex,semithick] (c2) arc[radius=9.7, start angle=-40, end angle=-3] node (bb2) {};
  \draw[<-,>=latex,semithick] (c3) arc[radius=9.7, start angle=-20, end angle=30] node (a33) {};
  \draw (c3) arc[radius=9.7, start angle=-20, end angle=15] node (b3) {$\prescript{b_3}{}\;\bullet_{\;\;\;\;}$};
  \draw[-<,>=latex,semithick] (c3) arc[radius=9.7, start angle=-20, end angle=17] node (bb3) {};
  \draw[<-,>=latex,semithick] (cl) arc[radius=9.7, start angle=10, end angle=60] node (all) {};
  \draw[-<,>=latex,semithick] (cl) arc[radius=9.7, start angle=10, end angle=47] node (bl) {$\prescript{b_{k}}{}\;\bullet_{\;\;\;\;}$};
  \draw[-<,>=latex,semithick] (cl) arc[radius=9.7, start angle=10, end angle=49] node (bbl) {};	
  
\end{tikzpicture}
		\caption{The $(n,k)$-arch process \label{fig:arch_proc}}
	\end{center}
\end{wrapfigure}
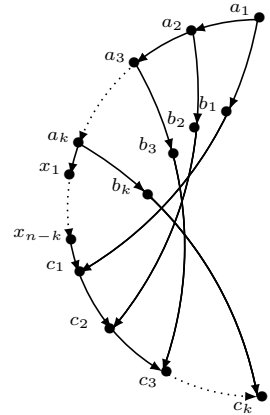

In the present work we focus on processes built on \emph{synchronization}
by means of \emph{futures} (or \emph{promises}). 
This concept is an abstraction 
for processes based on a main execution thread but
allowing to delay some computations.
These computations are run asynchronously and are represented
by an object that can be queried in two ways: \texttt{finish?}
to know if the computation has terminated and
\texttt{get} to retrieve the result of the computation (and properly
proceed the synchronization).

The structure of a general concurrent process with futures is
more or less a directed acyclic graph. Since the quantitative study
of such increasingly labeled graphs seems out of reach, we
restrict ourselves to the study of \emph{arch processes}, a simplistic
model of processes with futures. An arch process is composed of a main
\emph{trunk} from which start several \emph{arches} (modelizing
futures). The general shape of such a process is given in
Fig.~\ref{fig:arch_proc}. The arch processes are based on two
parameters related to their sizes and their numbers of arches. 
To our knowledge the increasingly labeled structures are not specifiable in the
sense of \cite{FS09}.

For this limited model we exhibit an \emph{exact} and an \emph{asymptotic formula} for the number of
increasing labelings.
The second main contribution of this paper is composed of two algorithms.
The first one is an uniform random sampler for runs of a given arch process
and the second one is an \emph{unranking}
algorithm which allows to obtain an exhaustive builder of runs.

The paper is organized as follows. The next section is devoted to the
formal description of $(n,k)$-arch processes and gives the solution of the recurrence
equation driving their numbers of runs. In
Section~\ref{sec:alg} we prove the algebraicity of the bivariate generating
function, we give a closed form formula for it and,
we give the asymptotic behaviors of the diagonal coefficients 
of the functions. Section~\ref{sec:algo} carefully describes both algorithms.

	\section{The arch processes and their runs}
	

\begin{definition}
	Let $n$ and $k$ be two positive integers with $k \leq n+1$.
	The $(n,k)$-\emph{arch process}, denoted by~$A_{n,k}$, is built in the following way:
	\begin{itemize}
		\item the trunk of the process: a sequence of $n+k$ actions $a_1, a_2, \dots, a_k, x_1, x_2, \dots, x_{n-k}, c_1, c_2, \dots, c_k$
			represented in Fig.~\ref{fig:arch_proc} on a semicircle;
		\item for all $i\in\{1, \dots, k\}$, the actions $a_i$ and $c_i$ are directly linked by an arch containing a single action~$b_i$.
	\end{itemize}
\end{definition}
We remark the value $k$ corresponds to the number of arches in the process, and $n$ is the length (in the trunk)
between both extremities of each arch.
There are two extreme cases: when $k=n$, it corresponds to the arch processes that do not contain any node $x_i$ in the trunk,
and the case $k=n+1$ that corresponds to the case where both the nodes $a_k$ and $c_1$ are merged into a single node
(and thus there is no node $x_i$).

In Fig.~\ref{fig:arch_proc} representing the $(n,k)$-arch process,
the precedence constraints are encoded with the directed edges such that $a \rightarrow b$ means
that the action $a$ precedes $b$.
We remark that the $(n,k)$-arch process contains exactly $n+2k$ actions.

Due to the intertwining of the arches, we immediately observe when $k$ is
larger than $1$ then the arch processes are not Series-Parallel processes.
Hence the results we exhibited in our papers~\cite{BDGP17bis,BDGP17}
cannot be applied in this context.

\begin{wrapfigure}[15]{r}{3.2cm}
\begin{tikzpicture}[xscale=0.4, yscale=0.4]
	\node (a1) at (0,5) {\small $\prescript{\textcolor{red}{1\;}}{\phantom{f}}\bullet_{\;\;\;}$};
	\draw[->,>=latex,semithick] (a1) arc[radius=5, start angle=90, end angle=113] node (a2) {\small $\prescript{\textcolor{red}{\:\:3\;}}{}\bullet_{\;\;\;\;}$};
	\draw[->,>=latex,semithick] (a2) arc[radius=5, start angle=113, end angle=136] node (a3) {\small $\prescript{\textcolor{red}{\;\;4}}{}\bullet_{\;\;\;\;}$};
	\draw[->,>=latex,semithick] (a3) arc[radius=5, start angle=136, end angle=159] node (al) {\small $\prescript{\textcolor{red}{\;\;6}}{}\bullet_{\;\;\;\;}$};
	\draw[->,>=latex,semithick] (al) arc[radius=5, start angle=159, end angle=180] node (x1) {\small $\prescript{\textcolor{red}{\;\;7}}{}\bullet_{\;\;\;\;}$};
	\draw[->,>=latex,semithick] (x1) arc[radius=5, start angle=180, end angle=201] node (c1) {\small $\prescript{\textcolor{red}{\;9\;}}{}\bullet_{\;\;\;\;}$};
	\draw[->,>=latex,semithick] (c1) arc[radius=5, start angle=201, end angle=224] node (c2) {\small $\prescript{\textcolor{red}{11} \;\;}{}\bullet_{\;\;\;\;\;\;}$};
	\draw[->,>=latex,semithick] (c2) arc[radius=5, start angle=224, end angle=247] node (c3) {\small $\prescript{\phantom{f}}{\;\textcolor{red}{12}}\bullet_{\;\;\;\;\;}$};
	\draw[->,>=latex,semithick] (c3) arc[radius=5, start angle=247, end angle=270] node (cl) {\small $\prescript{\phantom{f}}{\textcolor{red}{13}}\bullet_{\;\;\;}$};

	\draw[<-,>=latex,semithick] (c1) arc[radius=9.7, start angle=-60, end angle=-10] node (a11) {};
	\draw (c1) arc[radius=9.7, start angle=-60, end angle=-20] node (b1) {\small $\prescript{\textcolor{red}{2\:}}{}\bullet_{\;\;\;}$};	
	\draw[-<,>=latex,semithick] (c1) arc[radius=9.7, start angle=-60, end angle=-18] node (bb1) {};
	\draw[<-,>=latex,semithick] (c2) arc[radius=9.7, start angle=-37, end angle=13] node (a22) {};
	\draw (c2) arc[radius=9.7, start angle=-37, end angle=3] node (b2) {\small $\prescript{\textcolor{red}{10\;}}{}\bullet_{\;\;\;}$};
	\draw[-<,>=latex,semithick] (c2) arc[radius=9.7, start angle=-37, end angle=5] node (bb2) {};
	\draw[<-,>=latex,semithick] (c3) arc[radius=9.7, start angle=-14, end angle=36] node (a33) {};
	\draw (c3) arc[radius=9.7, start angle=-14, end angle=26] node (b3) {\small $\prescript{\textcolor{red}{\;5}}{}\;\bullet_{\;\;\;\;}$};
	\draw[-<,>=latex,semithick] (c3) arc[radius=9.7, start angle=-14, end angle=28] node (bb3) {};
	\draw[<-,>=latex,semithick] (cl) arc[radius=9.7, start angle=10, end angle=60] node (all) {};
	\draw (cl) arc[radius=9.7, start angle=10, end angle=47] node (bl) {\small $\prescript{}{\textcolor{red}{8}}\;\bullet_{\;\;\;\;\;}$};
	\draw[-<,>=latex,semithick] (cl) arc[radius=9.7, start angle=10, end angle=50] node (bbl) {};
			
		\end{tikzpicture}
	\caption{A run of the $(5,4)$-arch process
		\label{fig:run}}
\end{wrapfigure}
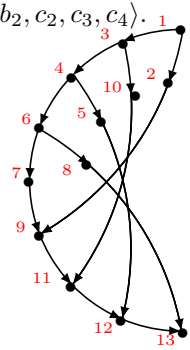

\begin{definition}
	For a concurrent process a \emph{run} is a total order of the actions that is compatible with 
	the precedence constraints describing the process.
\end{definition}

\begin{definition}
	An \emph{increasing labeling} for a concurrent process containing $\ell$ actions
	is a bijection between the integers $\{1, \dots, \ell\}$
	and the actions of the process, satisfying the following constraint: 
	if an action $a$ precedes an action $b$
	then the label associated to $a$ is smaller than the one related to $b$.
\end{definition}

In Fig.~\ref{fig:run} we have represented an increasing labeling 
of the $(5,4)$-arch process $A_{5,4}$ corresponding to the run
$\langle a_1, b_1, a_2, a_3, b_3, a_4, x_1, b_4, c_1, b_2, c_2, c_3, c_4 \rangle$.
As one can see, every directed path (induced by the precedence relation) is increasingly labeled.
Our quantitative goal is to calculate the number of runs for a given arch process.

\begin{proposition}
	The number of runs of a concurrent process
	is the number of increasing labelings of the actions of the process.
\end{proposition}

While there is the classical hook-length formula for tree-processes~\cite{Knuth98,BGP16}
and its generalization for Series-Parallel processes~\cite{BDGP17bis},
to the best of our knowledge, no closed form formula is known for more general classes of processes.
In the rest of the paper, for a given process $A$, we denote by $\sigma(A)$ its
number of runs.

\begin{figure}[htb]
	\begin{center}
		\begin{tikzpicture}[xscale=0.33, yscale=0.33]
			
	\node (a1) at (0,5) {\small $\prescript{a_1}{\phantom{f}}\bullet_{\;\;\;}$};
	\draw[->,>=latex,semithick] (a1) arc[radius=5, start angle=90, end angle=110] node (a2) {\small $\prescript{a_2}{}\bullet_{\;\;\;\;}$};
	\draw[->,>=latex,semithick] (a2) arc[radius=5, start angle=110, end angle=130] node (a3) {\small $\prescript{a_3}{}\bullet_{\;\;\;\;}$};
	\draw[dotted,->,>=latex,semithick] (a3) arc[radius=5, start angle=130, end angle=160] node (al) {\small $\prescript{a_k}{}\bullet_{\;\;\;\;}$};
	\draw[->,>=latex,semithick] (al) arc[radius=5, start angle=160, end angle=170] node (x1) {\small $\prescript{x_1}{}\bullet_{\;\;\;\;}$};
	\draw[dotted,->,>=latex,semithick] (x1) arc[radius=5, start angle=170, end angle=190] node (xr) {\small $\prescript{x_{n-k}}{}\bullet_{\;\;\;\;\;\;\;\;}$};
	\draw[->,>=latex,semithick] (xr) arc[radius=5, start angle=190, end angle=200] node (c1) {\small $\prescript{c_1 \;}{}\bullet_{\;\;\;\;}$};
	\draw[->,>=latex,semithick] (c1) arc[radius=5, start angle=200, end angle=220] node (c2) {\small $\prescript{c_2 \;\;}{}\bullet_{\;\;\;\;\;\;}$};
	\draw[->,>=latex,semithick] (c2) arc[radius=5, start angle=220, end angle=240] node (c3) {\small $\prescript{\phantom{f}}{c_3}\bullet_{\;\;}$};
	\draw[dotted,->,>=latex,semithick] (c3) arc[radius=5, start angle=240, end angle=270] node (cl) {\small $\prescript{\phantom{f}}{c_k}\bullet_{\;\;}$};
	\draw[->,>=latex,semithick] (cl) arc[radius=5, start angle=270, end angle=290] node (cc1) {\small $\prescript{\phantom{f}\;\;\;}{\;\;}\bullet_{c'_1}$};

	\draw[<-,>=latex,semithick] (c1) arc[radius=9.7, start angle=-60, end angle=-10] node (a11) {};
	\draw (c1) arc[radius=9.7, start angle=-60, end angle=-20] node (b1) {\small $\prescript{b_1}{}\bullet_{\;\;\;}$};	
	\draw[-<,>=latex,semithick] (c1) arc[radius=9.7, start angle=-60, end angle=-18] node (bb1) {};
	\draw[<-,>=latex,semithick] (c2) arc[radius=9.7, start angle=-40, end angle=10] node (a22) {};
	\draw (c2) arc[radius=9.7, start angle=-40, end angle=-2] node (b2) {\small $\prescript{b_2}{}\bullet_{\;\;\;}$};
	\draw[-<,>=latex,semithick] (c2) arc[radius=9.7, start angle=-40, end angle=0] node (bb2) {};
	\draw[<-,>=latex,semithick] (c3) arc[radius=9.7, start angle=-20, end angle=30] node (a33) {};
	\draw (c3) arc[radius=9.7, start angle=-20, end angle=15] node (b3) {\small $\prescript{b_3}{}\;\bullet_{\;\;\;\;}$};
	\draw[-<,>=latex,semithick] (c3) arc[radius=9.7, start angle=-20, end angle=17] node (bb3) {};
	\draw[<-,>=latex,semithick] (cl) arc[radius=9.7, start angle=10, end angle=60] node (all) {};
	\draw[-<,>=latex,semithick] (cl) arc[radius=9.7, start angle=10, end angle=47] node (bl) {\small $\prescript{b_{k}}{}\;\bullet_{\;\;\;\;}$};
	\draw[-<,>=latex,semithick] (cl) arc[radius=9.7, start angle=10, end angle=49] node (bbl) {};	

		\end{tikzpicture}
		\begin{tikzpicture}[xscale=0.33, yscale=0.33]
			
	\node (a1) at (0,5) {\small $\prescript{a_1}{\phantom{f}}\bullet_{\;\;\;}$};
	\draw[->,>=latex,semithick] (a1) arc[radius=5, start angle=90, end angle=110] node (a2) {\small $\prescript{a_2}{}\bullet_{\;\;\;\;}$};
	\draw[->,>=latex,semithick] (a2) arc[radius=5, start angle=110, end angle=130] node (a3) {\small $\prescript{a_3}{}\bullet_{\;\;\;\;}$};
	\draw[dotted,->,>=latex,semithick] (a3) arc[radius=5, start angle=130, end angle=160] node (al) {\small $\prescript{a_k}{}\bullet_{\;\;\;\;}$};
	\draw[->,>=latex,semithick] (al) arc[radius=5, start angle=160, end angle=170] node (x1) {\small $\prescript{x_1}{}\bullet_{\;\;\;\;}$};
	\draw[dotted,->,>=latex,semithick] (x1) arc[radius=5, start angle=170, end angle=190] node (xr) {\small $\prescript{x_{n-k}}{}\bullet_{\;\;\;\;\;\;\;\;}$};
	\draw[->,>=latex,semithick] (xr) arc[radius=5, start angle=190, end angle=200] node (c1) {\small $\prescript{c_1 \;}{}\bullet_{\;\;\;\;}$};
	\draw[->,>=latex,semithick] (c1) arc[radius=5, start angle=200, end angle=220] node (c2) {\small $\prescript{c_2 \;\;}{}\bullet_{\;\;\;\;\;\;}$};
	\draw[->,>=latex,semithick] (c2) arc[radius=5, start angle=220, end angle=240] node (c3) {\small $\prescript{\phantom{f}}{c_3}\bullet_{\;\;}$};
	\draw[dotted,->,>=latex,semithick] (c3) arc[radius=5, start angle=240, end angle=270] node (cl) {\small $\prescript{\phantom{f}}{c_k}\bullet_{\;\;}$};
	\draw[->,>=latex,semithick] (cl) arc[radius=5, start angle=270, end angle=290] node (cc1) {\small $\prescript{\phantom{f}\;\;\;}{\;\;}\bullet_{c'_1}$};
	
	\draw[<-,>=latex,semithick] (cc1) arc[radius=7.9, start angle=-29, end angle=48] node (a11) {};
	\draw (cc1) arc[radius=7.9, start angle=-29, end angle=25] node (b1) {\small $\prescript{b_1}{}\bullet_{\;\;\;}$};	
	\draw[-<,>=latex,semithick] (cc1) arc[radius=7.9, start angle=-29, end angle=27] node (bb1) {};
	\draw[<-,>=latex,semithick] (c2) arc[radius=9.7, start angle=-40, end angle=10] node (a22) {};
	\draw (c2) arc[radius=9.7, start angle=-40, end angle=-2] node (b2) {\small $\prescript{b_2}{}\bullet_{\;\;\;}$};
	\draw[-<,>=latex,semithick] (c2) arc[radius=9.7, start angle=-40, end angle=0] node (bb2) {};
	\draw[<-,>=latex,semithick] (c3) arc[radius=9.7, start angle=-20, end angle=30] node (a33) {};
	\draw (c3) arc[radius=9.7, start angle=-20, end angle=15] node (b3) {\small $\prescript{b_3}{}\;\bullet_{\;\;\;\;}$};
	\draw[-<,>=latex,semithick] (c3) arc[radius=9.7, start angle=-20, end angle=17] node (bb3) {};
	\draw[<-,>=latex,semithick] (cl) arc[radius=9.7, start angle=10, end angle=60] node (all) {};
	\draw[-<,>=latex,semithick] (cl) arc[radius=9.7, start angle=10, end angle=47] node (bl) {\small $\prescript{b_{k}}{}\;\bullet_{\;\;\;\;}$};
	\draw[-<,>=latex,semithick] (cl) arc[radius=9.7, start angle=10, end angle=49] node (bbl) {};	
	
		\end{tikzpicture}
		\begin{tikzpicture}[xscale=0.33, yscale=0.33]
			
	\node (a1) at (0,5) {\small $\prescript{a_1}{\phantom{f}}\bullet_{\;\;\;}$};
	\draw[->,>=latex,semithick] (a1) arc[radius=5, start angle=90, end angle=110] node (a2) {\small $\prescript{a_2}{}\bullet_{\;\;\;\;}$};
	\draw[->,>=latex,semithick] (a2) arc[radius=5, start angle=110, end angle=130] node (a3) {\small $\prescript{a_3}{}\bullet_{\;\;\;\;}$};
	\draw[dotted,->,>=latex,semithick] (a3) arc[radius=5, start angle=130, end angle=160] node (al) {\small $\prescript{a_k}{}\bullet_{\;\;\;\;}$};
	\draw[->,>=latex,semithick] (al) arc[radius=5, start angle=160, end angle=170] node (x1) {\small $\prescript{x_1}{}\bullet_{\;\;\;\;}$};
	\draw[dotted,->,>=latex,semithick] (x1) arc[radius=5, start angle=170, end angle=190] node (xr) {\small $\prescript{x_{n-k}}{}\bullet_{\;\;\;\;\;\;\;\;}$};
	\draw[->,>=latex,semithick] (xr) arc[radius=5, start angle=190, end angle=200] node (c1) {\small $\prescript{c_1 \;}{}\bullet_{\;\;\;\;}$};
	\draw[->,>=latex,semithick] (c1) arc[radius=5, start angle=200, end angle=220] node (c2) {\small $\prescript{c_2 \;\;}{}\bullet_{\;\;\;\;\;\;}$};
	\draw[->,>=latex,semithick] (c2) arc[radius=5, start angle=220, end angle=240] node (c3) {\small $\prescript{\phantom{f}}{c_3}\bullet_{\;\;}$};
	\draw[dotted,->,>=latex,semithick] (c3) arc[radius=5, start angle=240, end angle=270] node (cl) {\small $\prescript{\phantom{f}}{c_k}\bullet_{\;\;}$};
	\draw[->,>=latex,semithick] (cl) arc[radius=5, start angle=270, end angle=290] node (cc1) {\small $\prescript{\phantom{f}\;\;\;}{\;\;}\bullet_{c'_1}$};
	
	\draw[<-,>=latex,semithick] (cc1) arc[radius=9.7, start angle=24, end angle=77] node (a11) {};
	\draw (cc1) arc[radius=9.7, start angle=24, end angle=45] node (b1) {\small $\prescript{\phantom{f}\;\;\;}{\;\;}\bullet_{\;b_1}$};	
	\draw[-<,>=latex,semithick] (cc1) arc[radius=9.7, start angle=24, end angle=47] node (bb1) {};
	\draw[<-,>=latex,semithick] (c2) arc[radius=9.7, start angle=-40, end angle=10] node (a22) {};
	\draw (c2) arc[radius=9.7, start angle=-40, end angle=-2] node (b2) {\small $\prescript{b_2}{}\bullet_{\;\;\;}$};
	\draw[-<,>=latex,semithick] (c2) arc[radius=9.7, start angle=-40, end angle=0] node (bb2) {};
	\draw[<-,>=latex,semithick] (c3) arc[radius=9.7, start angle=-20, end angle=30] node (a33) {};
	\draw (c3) arc[radius=9.7, start angle=-20, end angle=15] node (b3) {\small $\prescript{b_3}{}\;\bullet_{\;\;\;\;}$};
	\draw[-<,>=latex,semithick] (c3) arc[radius=9.7, start angle=-20, end angle=17] node (bb3) {};
	\draw[<-,>=latex,semithick] (cl) arc[radius=9.7, start angle=10, end angle=60] node (all) {};
	\draw[-<,>=latex,semithick] (cl) arc[radius=9.7, start angle=10, end angle=47] node (bl) {\small $\prescript{b_{k}}{}\;\bullet_{\;\;\;\;}$};
	\draw[-<,>=latex,semithick] (cl) arc[radius=9.7, start angle=10, end angle=49] node (bbl) {};	

		\end{tikzpicture}
		\begin{tikzpicture}[xscale=0.33, yscale=0.33]
			
	\node (a1) at (0,5) {\small $\prescript{a_1}{\phantom{f}}\bullet_{\;\;\;}$};
	\draw[->,>=latex,semithick] (a1) arc[radius=5, start angle=90, end angle=110] node (a2) {\small $\prescript{a_2}{}\bullet_{\;\;\;\;}$};
	\draw[->,>=latex,semithick] (a2) arc[radius=5, start angle=110, end angle=130] node (a3) {\small $\prescript{a_3}{}\bullet_{\;\;\;\;}$};
	\draw[dotted,->,>=latex,semithick] (a3) arc[radius=5, start angle=130, end angle=160] node (al) {\small $\prescript{a_k}{}\bullet_{\;\;\;\;}$};
	\draw[->,>=latex,semithick] (al) arc[radius=5, start angle=160, end angle=170] node (x1) {\small $\prescript{x_1}{}\bullet_{\;\;\;\;}$};
	\draw[dotted,->,>=latex,semithick] (x1) arc[radius=5, start angle=170, end angle=190] node (xr) {\small $\prescript{x_{n-k}}{}\bullet_{\;\;\;\;\;\;\;\;}$};
	\draw[->,>=latex,semithick] (xr) arc[radius=5, start angle=190, end angle=200] node (c1) {\small $\prescript{c_1 \;}{}\bullet_{\;\;\;\;}$};
	\draw[->,>=latex,semithick] (c1) arc[radius=5, start angle=200, end angle=220] node (c2) {\small $\prescript{c_2 \;\;}{}\bullet_{\;\;\;\;\;\;}$};
	\draw[->,>=latex,semithick] (c2) arc[radius=5, start angle=220, end angle=240] node (c3) {\small $\prescript{\phantom{f}}{c_3}\bullet_{\;\;}$};
	\draw[dotted,->,>=latex,semithick] (c3) arc[radius=5, start angle=240, end angle=270] node (cl) {\small $\prescript{\phantom{f}}{c_k}\bullet_{\;\;}$};
	\draw[->,>=latex,semithick] (cl) arc[radius=5, start angle=270, end angle=290] node (cc1) {\small $\prescript{\phantom{f}\;\;\;}{\;\;}\bullet_{c'_1}$};

	\draw[<-,>=latex,semithick] (c1) arc[radius=2.3, start angle=-30, end angle=35] node (a11) {};
	\draw (c1) arc[radius=2.3, start angle=-30, end angle=10] node (b1) {\small $\prescript{\phantom{f}\;\;\;}{\;\;}\bullet_{\;b_1}$};	
	\draw[-<,>=latex,semithick] (c1) arc[radius=2.3, start angle=-30, end angle=18] node (bb1) {};
	\draw[<-,>=latex,semithick] (c2) arc[radius=9.7, start angle=-40, end angle=10] node (a22) {};
	\draw (c2) arc[radius=9.7, start angle=-40, end angle=-2] node (b2) {\small $\prescript{b_2}{}\bullet_{\;\;\;}$};
	\draw[-<,>=latex,semithick] (c2) arc[radius=9.7, start angle=-40, end angle=0] node (bb2) {};
	\draw[<-,>=latex,semithick] (c3) arc[radius=9.7, start angle=-20, end angle=30] node (a33) {};
	\draw (c3) arc[radius=9.7, start angle=-20, end angle=15] node (b3) {\small $\prescript{b_3}{}\;\bullet_{\;\;\;\;}$};
	\draw[-<,>=latex,semithick] (c3) arc[radius=9.7, start angle=-20, end angle=17] node (bb3) {};
	\draw[<-,>=latex,semithick] (cl) arc[radius=9.7, start angle=10, end angle=60] node (all) {};
	\draw[-<,>=latex,semithick] (cl) arc[radius=9.7, start angle=10, end angle=47] node (bl) {\small $\prescript{b_{k}}{}\;\bullet_{\;\;\;\;}$};
	\draw[-<,>=latex,semithick] (cl) arc[radius=9.7, start angle=10, end angle=49] node (bbl) {};	

		\end{tikzpicture}

	\caption{From left to right, the processes denoted $D_{n,k}, \overline{D}_{n,k}, \overline{D}_{n,k}^1$ and $\overline{D}_{n,k}^2$
		\label{fig:recurrence}}
	\end{center}
\end{figure}
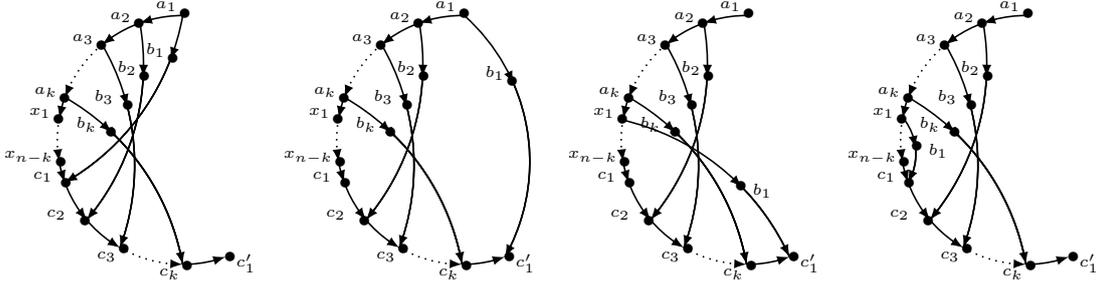

First, let us easily exhibit a lower bound and an upper bound (in the case $k < n+1$)
in order to obtain a first idea for the growth of the numbers of runs for the arch processes.
Remark that a similar approach could be used for the case when $k = n+1$.
We first enumerate the runs where all the nodes $b_i$'s are preceded by $a_k$,
and all of them precede the node $c_1$. This imposes new precedence constraints for the process,
and thus its number of runs is a lower bound for the total number of runs.
In this case the $b_i$'s permute without any constraint, i.e. $k!$ possibilities
and then each permutation of the $b_i$'s shuffles with the sequence $x_1, \dots, x_{n-k}$.
Thus we get the following lower bounds for the number of runs of $A_{n,k}$:
\[
	\sigma(A_{n,k}) \geq k! \binom{k + n-k }{k} = \frac{n!}{(n-k)!}.
\]
We now focus on an upper bound for the number of runs of $A_{n,k}$.
Here again we suppose that all the permutations of the $b_i$'s are possible,
but we allow each $b_i$ to appear everywhere between $a_1$ and $c_k$.
This constraint is satisfied by all the runs, but some
possibilities are not valid runs: thus we are computing an upper bound.
Once the permutation of the $b_i$'s is calculated, we shuffle it into the trunk (containing $n+k$ nodes):
\[
	\sigma(A_{n,k}) \leq k! \binom{k + n+k -1}{n+k -1} = \frac{(n+2k-1)!}{(n+k-1)!}.
\]
A refinement of these ideas for the bounds computation allows to exhibit a
recurrence formula for the value $\sigma(A_{n,k})$.
\begin{theorem}\label{thm:rec}
	Let $n$ and $k$ be two integers such that $0 \leq k \leq n+1$.
	The number $\sigma(A_{n,k})$ of runs of the process $A_{n,k}$
        is equal to $t_{n,k}$ with:
        \begin{equation}\label{eq:recurrence}
          t_{n,k}=\frac{n+2k-1}{2}t_{n, k-1} + \frac{n-k}{2}t_{n+1, k-1} \quad \text{ and } \quad t_{n,0}=1.
        \end{equation}

\end{theorem}
In order to provide the proof, we first introduce the four processes in Fig.\ref{fig:recurrence}.
Notice that they are not arch processes.
From the left handside to right handside,
the first process, denoted by $D_{n,k}$, is almost the process $A_{n,k}$. In fact, the single difference 
is that $D_{n,k}$ contains exactly one more action, denoted by $c'_1$, that is preceded by all the other actions.
The second process $\overline{D}_{n,k}$ is related to $D_{n,k}$ in the following way: the precedence relation starting at $b_1$
is replaced, instead of having $b_1 \rightarrow c_1$, it is $b_1 \rightarrow c'_1$.
Finally, for the two last processes $\overline{D}_{n,k}^1$ and $\overline{D}_{n,k}^2$, it is also the relations
$a_1 \rightarrow b_1 \rightarrow c_1$ which are modified.

\begin{proof}
The extreme case $A_{n,0}$ corresponds to a process without
any arch: just a trunk. Obviously it admits a single increasingly
labeling: it has a single run.
 
Suppose first that $k < n+1$.
The number $\sigma(A_{n,k})$ is equal to the number of runs $\sigma(D_{n,k})$ because for all runs,
the integer associated to $c'_1$ is 
inevitably the largest one: $2k+n+1$. Then we compute with some inclusion/exclusion rule the number $\sigma(D_{n,k})$:
\begin{equation}\label{eq:strcut}
	\sigma(D_{n,k}) = \sigma(\overline{D}_{n,k}) - \left( \sigma(\overline{D}^1_{n,k}) - \sigma(\overline{D}^2_{n,k}) \right).
\end{equation}
In fact we are focusing on the action preceded by $b_1$. In $D_{n,k}$ it corresponds to $c_1$. By modifying it by $c'_1$
in $\overline{D}_{n,k}$ we allow runs where $b_1$ appears after $c_1$, thus that are not valid for $D_{n,k}$.
We remove this number of non-valid runs with $\sigma(\overline{D}^1_{n,k}) - \sigma(\overline{D}^2_{n,k})$, by
playing with both actions $x_1$ and $c_1$.
To compute $\sigma(\overline{D}_{n,k})$, first omit the action $b_1$
(and its incoming and outgoing edges) ; the remaining process is a
$(n, k-1)$-arch process, up to renaming, with added top and bottom
actions ($a_1$ and $c'_1$) which do not modify the number of runs of
$A_{n,k-1}$. It remains to insert $b_1$ in this ``almost'' $A_{n,k-1}$,
somewhere between $a_1$ and $c'_1$: there is $(2\cdot(k-1)+n-1)+2 = 2k+n-1$
possibilities. The term $(2\cdot(k-1)+n-1)$ are the cases where $b_1$ is put between $a_2$ and $c_k$
and the term $2$ corresponds to the cases where $b_1$ is either before
$a_2$ or after $c_k$.
The process $\overline{D}^1_{n,k}$ is similar to the arch process $A_{n,k}$, there is only an
action $a_1$ that precedes it, so $\sigma(\overline{D}^1_{n,k}) =
t_{n,k}$.
Lastly, for the process $\sigma(\overline{D}^2_{n,k})$, forgetting
$b_1$ we recognize $A_{n+1,k-1}$ up to renaming, so  $b_1$ can be
inserted between $x_1$ and $c_1$: there are $n-k$ possibilities.
Finally we obtain the following equation
\[
	\sigma(A_{n,k}) = (n+2k-1) \cdot \sigma(A_{n,k-1}) - \sigma(A_{n,k}) + (n-k)\cdot \sigma(A_{n+1,k-1}).
\]
Induction principle let us conclude that equation~\eqref{eq:recurrence} is proved.

Suppose now that $k = n+1$. Here there is no action $x_i$ and both the nodes $a_k$
and $c_1$ are merged into a single node.
We can adapt equation~\eqref{eq:strcut} and obtain the same recurrence,
but via a small difference in the computation:
\[
	\sigma(A_{k-1,k}) = 3k \cdot \sigma(A_{k-1,k-1}) - \sigma(A_{k-1,k}) - \sigma(A_{k,k-1}).
\]
But since $k = n+1$, this recurrence is equal to equation~\eqref{eq:recurrence} too.
\end{proof}

Notice when $k > n+1$, for our model, it consists to merge the last actions $a_{n-i}$'s
with the first actions $c_i$'s. But the recursive formula~\eqref{eq:recurrence}
does not apply to such models: once $k>n+1$ the recurrence looses its
combinatorial meaning.\\
The next result exhibits a closed form formula for the number of runs of the
arch processes.
\begin{theorem}\label{theo:nb_processes}
	Let $n$ and $k$ be integers such that $0 < k \leq n+1$.
	The number\footnote{In Theorem~\ref{theo:nb_processes} we use the convention that the
	sum over the sequence of $i_j$'s is equal to 1 when $s=0$.}
	of runs of the $(n,k)$-arch process is
{\footnotesize
	\[
		\sigma(A_{n,k}) = \frac{(2k+n-1)!!}{2^{k-1}} \sum_{s=0}^{k-1} \frac{(n+s) \; \pari(n,s)}{(n+s+1)!!} \; 
			\sum_{1 \leq i_1 < i_2 < \dots < i_s \leq k \;} \;\; \prod_{j=1}^s (i_j+j+n-k-1) \frac{\Gamma\left(\frac{2k+n-2i_j+j}{2} +1 \right)}{\Gamma\left(\frac{2k+n-2i_j +j+1}{2} +1 \right)},
	\]
}
	with the following function 
	\[
		\pari(n,s) = \left\{ \begin{array}{l c l}
				\frac{1}{2^{s/2}} & & \mbox{if $s$ is even} \\[2pt]
				\frac{\sqrt{\pi}}{2^{(s+1)/2}} & & \mbox{if $s$ is odd and $n$ is even} \\[2pt]
				\frac{1}{2^{(s-1)/2} \sqrt{\pi}} & & \mbox{if $s$ is odd and $n$ is odd.}
			\end{array} \right.
	\]
\end{theorem}
Let us recall the double factorial notation: for $n\in \IN$, $n!! = n \cdot (n-2)!!$ with $0!! = 1!! = 1$.
We remark that the ratio of the two $\Gamma$-function is related to the
central binomial coefficient.
The asymptotic behavior of the sequence does not seem immediate to obtain
using this formula.

\begin{proof}[key-ideas]
	The formula for $\sigma(A_{n,k})$ is obtained by resolving the recurrence stated
	in equation~\eqref{eq:recurrence}. First remark that the calculation
	of $\sigma(A_{n,k})$ requires the values of $\sigma(A_{i,j})$ in the triangle
	such that $n \leq i \leq n+k$ and $0 \leq j \leq k-(i-n)$.
	The formula is computed by unrolling $k$ times the recurrence.
	In particular, the index~$s$ in the formula corresponds to the number
	of times we have used the second term of equation~\eqref{eq:recurrence}, 
	to reach the final term $\sigma(A_{n+s,0})$. The $i_j$'s values indicate in which iteration
	the second terms of equation~\eqref{eq:recurrence} have been chosen. They
	describe the path from $(n,k)$ to $(n+s,0)$.
	The brute formula obtained in this way is composed of a product of truncated double factorials
	that can be written as ratios of double factorial numbers.
	Finally, by coupling the adequate numerators and denominators in the product
	we exhibit several Wallis's ratios~\cite{AS64} that are easily simplified by using 	
	the $\Gamma$~function: 
	$ \displaystyle{\frac{(2n-1)!!}{(2n)!!} = \frac{1}{\sqrt{\pi}}\frac{\Gamma\left(n + \frac12\right)}{\Gamma\left(n+1\right)}}$.
\end{proof}

By using this closed form formula, or the bivariate recurrence (cf. equation~\eqref{eq:recurrence}),
we easily compute the first diagonals of the recurrence. The values of a given diagonal correspond to the class
of arch processes with the same number of actions $x_i$'s in the trunk.
{\footnotesize
\begin{align*}
	(\sigma(A_{k-1,k}))_{k\in \IN\setminus\{0,1\}} &= \left(1, 12, 170, 2940, 60760, 1466640, 40566680, 1266064800, 44030186200,1688858371200, \dots \right) \\
	(\sigma(A_{k,k}))_{k\in \IN^*} &= \left(1, 5, 44, 550, 8890, 176120, 4130000, 111856360, 3435632200, 117991273400, \dots \right) \\
	(\sigma(A_{k+1,k}))_{k\in \IN^*} &= \left(2, 11, 100, 1270, 20720, 413000, 9726640, 264279400, 8137329200, 280012733000, \dots \right) \\
	(\sigma(A_{k+2,k}))_{k\in \IN^*} &= \left(3, 19, 186, 2474, 41670, 850240, 20386800, 561863960, 17501627640, 608063465800, \dots \right) 
\end{align*}
}
We remark that the first terms of the sequence $(\sigma(A_{k+1,k}))_{k\in \IN^*}$ coincide with the first terms of
the sequence \texttt{A220433} (shifted by 2)
in OEIS\footnote{OEIS corresponds to the On-line Encyclopedia of Integer Sequences: \url{http://oeis.org/}.} . This sequence
is related to a specific Alia algebra and is exhibited in the paper of Khoroshkin and Piontkovski~\cite{KP15}.
In their paper, the exponential univariate generating function naturally appears as an algebraic function.
This motivates us to study in detail the bivariate generating function for $(t_{n,k})$ and
in particular its diagonals.

	\section{Algebraic generating functions}\label{sec:alg}

Let us associate to the bivariate sequence $(t_{n,k})_{n,k}$ the generating function, denoted by $A(z,u)$,
exponential in $u$ and ordinary in $z$: 
\[
	A(z,u)=\sum_{n\geq 0, k\geq 0} \frac{t_{n,k}}{k!} z^n u^k.
\]
Recall this series enumerates the increasing labelings of the arch processes, when $k\leq n+1$,
but has no combinatorial meaning beyond this bound.

\begin{proposition}\label{prop:holon}
	The bivariate generating function $A(z,u)$ is holonomic and satisfies the following 
	differential equation.
	\[
		\left( 2zu-2z-u \right) {\frac {\partial }{\partial u}}A \left( z,u \right) 
		+ \left( z-2 \right) A\left( z,u \right) +z \left( z+1 \right)
	 	{\frac {\partial }{\partial z}}A \left( z,u \right)	+ C(u) = 0.
	\]
\end{proposition}
The differential equation can be exhibited since the recursive 
behavior of $(t_{n,k})$ is not disturbed beyond the bound $k> n+1$.

\begin{proof}[key-ideas]
	The differential equation is directly obtained from the recurrence equation~\eqref{eq:recurrence}.
	The function $C(u)$ encodes the initial conditions of the equation.
	The differential equation satisfied by $A(z,u)$ ensures its holonomicity (cf. \cite{Stanley01,FS09}).
\end{proof}

It is important to remark that $C(u)$ is holonomic. In fact we have
$C(u)=u\frac {\partial }{\partial u} A(0,u)+2 A(0,u)$
and consequently $C(u)$ is holonomic as a specialization of an holonomic bivariate
generating function.
A direct computation for $C(u)$ exhibits the following differential equation
{\footnotesize
\begin{align*}
&4 \left( 24{u}^{2}+3u+1 \right) C(u) 
-4u\left( 84{u}^{2}-3u+1 \right) {\frac {\rm d}{{\rm d}u}} C(u) 
-2u^2\left( 216{u}^{2}-151u+13 \right) {\frac {{\rm d}^{2}}{{\rm d}{u}^{2}}} C(u) \\
& -2u^2 \left( 58{u}^{3}-75{u}^{2}+33u-2 \right) {\frac {{\rm d}^{3}}{{\rm d}{u}^{3}}} C(u) 
-u^3\left( 8{u}^{3}-15{u}^{2}+12u-4 \right) {\frac {{\rm d}^{4}}{{\rm d}{u}^{4}}} C(u) -8\left(3u+1\right)=0.
\end{align*}
}
Note that we prove also that $C(u)$ is solution of an algebraic equation.
This fact is really not obvious from a combinatorial point of view.
But it is deduced through the fact that the function $A(0,u)$ is algebraic:
{\footnotesize
\begin{equation}\label{eq:A(0,u)_alg}
	(8u^3-15u^2+12u-4) A(0,u)^3 + (12u^2-12u+6) A(0,u) -2u^3=0.
\end{equation}
}
The equation is obtained by a \emph{guess and prove approach}. Once it has been guessed
it remains to prove it by using the holonomic equation proven in Proposition~\ref{prop:holon}.
Thus we get
{\footnotesize
\begin{align*}
	\left( 8{u}^{3}-15{u}^{2}+12\,u-4 \right)^{3} {C(u)}^{3}+48 \left( 36{u}^{6}-120{u}^{5}+202{u}^{4}-199{u}^{3}+123{u}^{2}-44u+8 \right)  \left( u-1 \right)^{2} C(u) &\\
	+ 32 \left( 9{u}^{2}-12u+8 \right)  \left( u-1 \right)^{3} &=0.
\end{align*}
}
\begin{theorem}
	The function $A(z,u)$ is an algebraic function in ($z$ and $u$)
	whose annihilating polynomial has degree~$3$:
{\footnotesize
\begin{align*}
\left( 8u^3-15u^2+12u-4 \right) \left( z^3+6zu+3z^2-3z-1 \right) A(z,u)^3 +6z^2 \left( 8u^3-15u^2+12u-4 \right) A(z,u)^2 & \\
 +6 \left( 12z{u}^{3}-18z{u}^{2}-2{u}^{2}+13zu+2u-3z-1 \right) A(z,u) +2 &=0.
\end{align*}
}
\end{theorem}
Note that the choice to use a doubly exponential generating function
(in $u$ and $z$) for $(t_{n,k})$ would have make sense and would be
holonomic too (closure property of Borel transform). But it would not
be algebraic because the inappropriate asymptotic expansion
(cf. Theorem~\ref{theo:asympt}).


\begin{proof}
The fact that the initial conditions and a diagonal of $A(z,u)$ are algebraic suggests that it
could also be algebraic as a function of $z$ and $u$.
Applying a bivariate guessing procedure, we observe that bivariate function $H(z,u)=(u+1)(z^3+3z^2+6zu-3z-1) A(z,u)$
is such that $[z^n]H(z,u)=0$ for $n > 2$.
Furthermore $[z^j]H(z,u)$ is algebraic for $j=\{0,1,2\}$. So, let us calculate these $z$-extractions.
First recall that $[z^0]A(z,u)$ satisfies the algebraic equation~\eqref{eq:A(0,u)_alg}.
In the same vein, $[z^1]A(z,u)$ verifies the algebraic equation
{\footnotesize
\begin{align*}
	\left(8u^3-15u^2+12u-4\right)f(u)^3 + 3\left(8u^3-15u^2+12u-4\right)f(u)^2 &\\
	+ 3\left(8u^3-15u^2+10u-2\right)f(u) +8u^3 -15u^2+6u&=0,
\end{align*}
}
and finally $[z^2]A(z,u)$ verifies the algebraic equation
{\footnotesize
\begin{align*}
	\left( 8{u}^{3}-15u^{2}+12u-4 \right) {f(u)}^{3}+ \left( -24{u}^{3}+45{u}^{2}-36u+12 \right) {f(u)}^{2} &\\
	+ \left( -72{u}^{3}+135{u}^{2}-84u+18 \right) f(u) -40{u}^{3}+75{u}^{2}-36u &=0.
\end{align*}
}
Thus we obtain
\begin{align*}
	[z^0] H(z,u) &= -(1+u) A(0,u) \\
	[z^1] H(z,u) &= -1 +(u+1) \left((6u-3)A(0,u)-[z^1]A(z,u) \right) \\
	[z^2] H(z,u) &=  \left( u+1 \right) \left((6u-3) [z^1]A(z,u) - [z^2]A(z,u) +3 A(0,u)+(6u-4) \right) .
\end{align*}
Finally we get
\[
	A(z,u)= \frac{[z^{\leq 2}]H(z,u)}{\left(u+1 \right) \left( z^3+3z^2+6uz-3z-1 \right)}.
\]
By using the elimination theory, we finally get a closed form algebraic equation for $A(z,u)$ of degree~$27$,
that obviously cannot fit in the conference paper format.
Nevertheless, this equation is not minimal. By simplifying it,
we finally get a minimal polynomial of degree~3 which annihilates $A(z,u)$:
{\footnotesize
\begin{align*}
	\left( 8u^3-15u^2+12u-4 \right) \left( z^3+3z^2+6zu-3z-1 \right) A(z,u)^3
	+6z^2 \left( 8u^3-15u^2+12u-4 \right) A(z,u)^2  & \\
	+6 \left( 12z{u}^{3}-18z{u}^{2}-2{u}^{2}+13zu+2u-3z-1 \right) A(z,u) +2 &=0.
\end{align*}
}
A direct proof by recurrence confirms the validity of this equation. 
\end{proof}

We remark in the previous section that the diagonals of the function $A(z,u)$
are of particular interest because they define subclasses of arch processes
with a fixed number of actions $x_i$'s covered by all the arches.
In order to extract the generating functions of these subclass, we could use
the Cauchy formula to compute $[u^0]A(z/u, u)$ and so on; we would keep the
holomicity property of the sequences but not their algebraicity. So,
we prefer to define the generating function $B(z,u) = A(z/u,u)$.
A similar proof that for the case $A(z,u)$ can be done to prove the
algebraicity of $B(z,u)$.
In particular, it exhibits the following algebraic equation satisfied by $B(z,u)$
{\footnotesize
\begin{align*}
	\left( 9{u}^{2}+12u-4 \right) \left( {z}^{3}+3{z}^{2}+6u-3z-1 \right) {B(z,u)}^{3} +  6 {z}^{2} \left( 9{u}^{2}+12u-4\right) {B(z,u)}^{2} & \\
	+ 6 \left( 18{u}^{2}z-18{u}^{2}+6uz+9u-3z-1 \right) B(z,u) + 2 \left(6u -1\right)^2 &=0.
\end{align*}	
}	
In particular, $B(0,u)$ is associated to the sequence $(t_{k,k})_k$, $[z^1]B(z,u)$ corresponds to the
sequence $(t_{k-1,k})_k$ and so on.
By specializing $z=0$ in the latter algebraic equation then by resolving it through the
Vi\`ete-Descartes approach for the resolution of cubic equation --detailed in the paper~\cite{Nickalls06}--,
we obtain the following closed form formula corresponding to the branch that is analytic in 0:
\[
	B(0,u) = \sqrt{2} \sqrt{\frac{1-3u}{1-3u-\frac94 u^2}} \cos \left( \frac13 
		\arccos \left( \frac{6u-1}{\sqrt{2}(1-3u)} \sqrt{\frac{1-3u-\frac94 u^2}{1-3u}} \right)  \right).
\]
Although the way we have represented $B(0,u)$ could suggest a singularity when the argument of the $\arccos$ function
is equal to $1$, the function admits an analytic continuation up to its dominant singularity~$\rho$, solution of 
$1-3u-\frac94 u^2=0$, thus corresponding to $\rho = \displaystyle{\frac{2}{3}\left(\sqrt{2}-1\right)}$.
Furthermore, by studying the global generating function $B(z,u)$, we obtain its singular expansion.
\begin{lemma}\label{lem:sing_exp}
	Near the singularity when $u$ tends to $\rho$, the function $B(z,u)$ satisfies
	\[
		B(z,u) \underset{u\rightarrow \rho}= a(z) + \frac{b(z)}{\sqrt{\rho - u}} + o\left( (\rho - u)^{-1/2}\right),
	\] 
	with $a(z)$ and $b(z)$ two functions independent from $u$.
\end{lemma}
By using this result we deduce the asymptotic behaviors of the diagonal coefficients of $A(z,u)$.
\begin{theorem}\label{theo:asympt}
Let $i$ be a given integer, and $k$ tend to infinity:
\[
	t_{k+i,k} \underset{k\rightarrow \infty}\sim \gamma_i \; \frac{\rho^{-k}}{\sqrt{k}} \; k! \qquad 
		\text{ with } \gamma_0 = \frac12 \sqrt{\frac{3}{\sqrt{2}\pi} \left(\sqrt{2}-1\right)} 
		 \quad \text{ and } \forall i\geq -1,   
		\gamma_i = \left(\frac{1}{\sqrt{2}-1}\right)^i \gamma_0.
\] 
\end{theorem}
This theorem is a direct consequence of Lemma~\ref{lem:sing_exp}. The 
$(\gamma_i)_i$ can be deduced by asymptotic matching.

Finally, by computing $[z^1]B(z,u)$ with the algebraic function it satisfies,
we prove that its second derivative is solution of the algebraic function exhibited in OEIS~\texttt{A220433}.

	\section{Uniform random generation of runs}\label{sec:algo}

We now introduce an algorithm to uniformly sample runs of a given arch process~$A_{n,k}$.
Our approach is based on the recursive equations~\eqref{eq:recurrence}
and~\eqref{eq:strcut} for the sequence $(t_{n,k})$.
Here we deal with the cases $k\leq n$ and avoid the limit case $k=n+1$.
Although the latter limit case satisfies this equation too, its proof
is based on an other combinatorial approach, and so the construction
of a run cannot be directly
deduced form the combinatorial approach proposed for the cases $k\leq
n$. Of course, a simple adaptation of the algorithm
presented below would allow to sample in $A_{k-1,k}$, but the lack of
space avoid us to present it here.

Our algorithm is a \emph{recursive generation algorithm}.
But since the objects are not specified in a classical Analytic Combinatorics's way,
we can not use the results of~\cite{FZVC94}.
As usual for recursive generation, the first step consists in the computation
and the memorization of the value $t_{n,k}$
and all the intermediate values $(t_{i,j})$ needed for the calculation of $t_{n,k}$.
\begin{proposition}\label{prop:pre_comput}
  In order to compute the value $t_{n,k}$, we need to calculate the values in the bi-dimensional set
  $\left\{t_{i,j} ~|~ n \leq i \leq n+k \text{ and } 0 \leq j \leq k-(i-n) \right\}$.
  This computation is done with $\BigO\left(k^2\right)$ arithmetic operations.
\end{proposition}
Recall that the coefficient computations are done only once for a given pair $(n,k)$,
and then many runs 
can be drawn uniformly for $A_{n,k}$ by using the recursive generation algorithm.

Let us present the way we exploit the recurrence
equation~\eqref{eq:strcut} to design the sampling method.
The main problem that we encounter is the presence of a minus sign in
the recurrence equation. Let us rewrite it in a slightly different way:
$\displaystyle{\sigma(D_{n,k}) + \sigma(\overline{D}^1_{n,k}) =  \sigma(\overline{D}_{n,k}) + \sigma(\overline{D}^2_{n,k})}$.

Recall that the structures under consideration are depicted in Fig.~\ref{fig:recurrence}.
We introduce the classes of increasingly labeled structures from 
$D_{n,k}, \overline{D}^1_{n,k}, \overline{D}_{n,k}$ and $\overline{D}^2_{n,k}$, respectively denoted 
by $I_{n,k}, \overline{I}^1_{n,k}, \overline{I}_{n,k}$ and $\overline{I}^2_{n,k}$.
Remark that the number of runs of $A_{n,k}$ is equal to $|I_{n,k}|$,
where the function $|\cdot|$ corresponds to the cardinality of the considered class.
Obviously the equation on the cardinalities can be written directly
on the classes
$I_{n,k} \cup \overline{I}^1_{n,k}  = \overline{I}_{n,k} \cup \overline{I}^2_{n,k}$
(since their intersections are empty: $I_{n,k}$ and $\overline{I}^1_{n,k}$
are distinct even if they are isomorphic).
Thus, we consider the problem of sampling the class $I_{n,k}\cup
\overline{I}^1_{n,k}$ where we bijectively replace the runs belonging
to $\overline{I}^1_{n,k}$ by ones of $I_{n,k}$ (which can be performed
recursively during the sampling procedure).
The Algorithm \textsc{Sampling}$(n,k)$ is based on the correspondence 
depicted in the Fig.~\ref{fig:recurrence} and its adaptation presented above on the
classes $I_{n,k} \cup \overline{I}^1_{n,k}$.
In each case the algorithm completes a recursively drawn run and apply some renaming on
the actions of that run. Then, it inserts the action $b_1$ according to the cases
$\overline{I}_{n,k} \backslash \overline{I}^1_{n,k}$, $\overline{I}^1_{n,k}$ or $\overline{I}^2_{n,k}$.
In the specific case $\overline{I}^1_{n,k}$, instead of $b_1$, it is 
the action $b_k$ that is inserted and the renaming
occurs in a similar fashion to obtain a run of $I_{n,k}$ from the one of $\overline{I}^1_{n,k}$.
\begin{theorem}
  The Algorithm \textsc{Sampling}$(n,k)$ builds uniformly at random a
  run of $A_{n,k}$ in $k$ recursive calls, once the coefficients computations and
  memorizations have been done.
\end{theorem}
Since each object of $I_{n,k}$
is sampled in two distinct ways, the uniform sampling in $I_{n,k}\cup \overline{I}^1_{n,k}$
induces the uniform sampling of $I_{n,k}$.

\begin{algorithm}[h]
	\caption{\label{algo:sampling} Uniform random sample for $I_{n,k}$}
	\begin{algorithmic}[1]
		\Function{Sampling}{$n,k$}
		\If{$k=0$}
		\State \textbf{return} $\langle x_1, x_2, \dots, x_n \rangle$
		\EndIf
		\State $r := \textsc{rand\_int}(0, 2\cdot t_{n,k}-1)$ \Comment an uniform integer between $0$ and $2\cdot t_{n,k}-1$ in $r$
		\If{$r < |\overline{I}_{n,k}|$} \Comment generation in $\overline{I}_{n,k}$
		
		\State $U := \textsc{Sampling} (n, k-1)$
		\State $p_{b} := 1 + r // t_{n,k-1}$ \Comment The position of the new $b$ to insert
		
		\If{$p_{b} > p_{x_1}$} \Comment generation in $\bar{I}^1_{n,k}$
		\State Rename $x_1$ by $a_k$ ; and each $x_i$ with $i > 1$ by $x_{i-1}$
		\State Insert $b_k$ at position $p_b$ ; and $c_k$ at the end of $U$
		
		\Else \Comment generation in $\overline{I}_{n,k} \backslash \bar{I}^1_{n,k}$
		\State In $U$, rename each $a_i$ (resp. $c_i$ and $b_i$) by $a_{i+1}$ (resp. $c_{i+1}$ and $b_{i+1}$)
		\State Rename $x_{n-k+1}$ by $c_1$
		\State Insert $b_1$ at position $p_b$ ; and $a_1$ at the head of $U$
		\EndIf
		\Else \Comment generation in $\overline{I}^2_{n,k}$
		\State $U := \textsc{Sampling}(n+1, k-1)$
		\State $p_b := 2 + (r- (n+2k-1)\cdot t_{n,k-1}) // t_{n+1,k-1}$
		\State Rename $x_{p_b}$ by $b_1$ and $x_{n-k+2}$ by $c_1$ ; and each $x_i$ with $i > p_b$ by $x_{i-1}$
		\State Insert $a_1$ at the head of $U$
		\EndIf
		\State \textbf{return} $U$
		\EndFunction
	\end{algorithmic}
	{\footnotesize
		Line $4$ and $17$ : the binary operator $//$ denotes the Euclidean division.\\
		The position of an action in a run is its arrival number (from 1 to the number of actions).
	}
\end{algorithm}
Focus on the run of $A_{5,4}$ depicted in Fig.~\ref{fig:run}: 
$\langle a_1, b_1, a_2, a_3, b_3, a_4, x_1, b_4, c_1, b_2, c_2, c_3, c_4 \rangle$.
It is either obtained from a (renamed) run of $\bar{I}^1_{5,4}$:
$\langle a_1, b_1, a_2, a_3, b_3, x_1, x_2, c_1, b_2, c_2, c_3
\rangle$ with $p_b = 8$ (Line 8 of the algorithm).
Or it is built from
$\langle a_1, a_2, b_2, a_3, x_1, b_3, x_2, b_1, c_1, c_2, c_3
\rangle$ of $\bar{I}_{5,4} \backslash \bar{I}^1_{5,4}$, with $p_b=1$ (Line 11).
But it cannot be built from a run of $\overline{I}^2_{5,4}$.

\begin{wrapfigure}[14]{r}{5.5cm}
	\includegraphics[width=5.5cm]{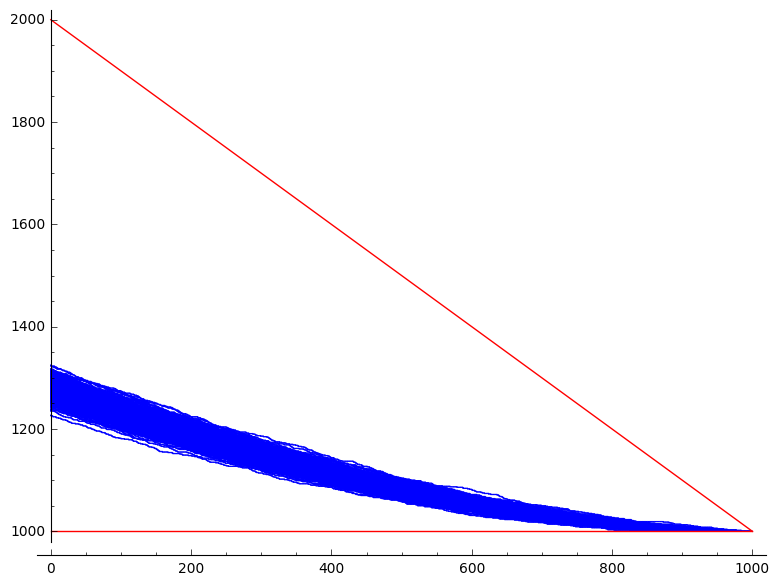}
	\caption{The terms $t_{i,j}$ needed for the sampling
		of $1000$ runs of $A_{1000,1000}$
		\label{fig:gen}}
\end{wrapfigure}
In Fig.~\ref{fig:gen},
we have uniformly sampled $1000$ runs for $A_{1000,1000}$ and we have
represented in blue points every pair $(k,n)$ corresponding
to an increasing sub-structure from $A_{n,k}$ that has been built
during the algorithm ($k$ for abscissa and $n$ for ordinate).
Only around $4.78\cdot 10^{4}$
sub-structures have been built among the $50\cdot 10^4$ 
inside the red lines which are calculated for the value $t_{1000,1000}$. 
At the beginning $n\approx k$ and
the \textbf{if} branch on Line 5 is preferred (instead of the \textbf{else} one on Line 15)
because the number of actions $x_i$'s is too small. After some 
recursive calls, the number of $x_i$'s actions has increased and then
both branches of the algorithm are taken with probabilities of the same order.
Recall that the constants $\gamma_i$'s (cf. Theorem~\ref{theo:asympt})
are evolving with an exponential growth.
Finally, we observe that only a small number of diagonals are necessary for the samplings.
Since the diagonals $(t_{n_i,k_i})$ for increasing sequences $(n_i)_i$ and $(k_i)_i$
follow P-recurrences (cf.~\cite{Lipshitz88}), a lazy calculations of the terms
of the necessary diagonals that envelop the blue points
(but that are much narrow to the blue points than both red lines)
would allow to minimize the pre-computations of Proposition~\ref{prop:pre_comput}.

We close this section with the presentation of an \emph{unranking} algorithm
for the construction of the runs of a given arch process~$A_{n,k}$.
This type of algorithm has been developed during the 70's by Nijenhuis and Wilf~\cite{NW75}
and introduced in the context of Analytic Combinatorics by Mart{\'i}nez and Molinero~\cite{MM03}.
Our algorithm is based on a bijection between the set of integers
$\{0, \dots, t_{n,k}-1\}$ and the set of runs of $A_{n,k}$. Here again
we restrict ourselves to the values $k\leq n$.
As usual for unranking algorithms, the first step consists in the computation
and the memorization of the values of a sequence.
But compared to the uniform random sampling, here we need more information
than the one given by the sequence $(t_{n,k})$.

To be able to reconstruct the run associated to a given rank, we need
to know the position
of the action $x_1$ in the recusively drawn run
in order to decide if the action $b_1$ appears before or after it.
First suppose $k < n$ and let $t_{n,k,\ell}$ be the number of runs in $\A_{n,k}$ whose
action $x_1$ appears at position $\ell$.
Let us denote by $I_{n,k,\ell}$ the associated combinatorial class.
We obtain directly a constructive recurrence for the sequence.
\begin{center}
$\displaystyle{	t_{n,k,\ell} = (\ell-2) \; t_{n,k-1,\ell-2} + (n-k) \; t_{n+1,k-1,\ell-1}
	 \quad \text{ and } \quad t_{n,0,1}=1; \; t_{n,0,\ell>1}=0}$.
\end{center}

\begin{proposition}\label{prop:pre_comput}
  The computation of $t_{n,k,\ell}$ is done with
  $\BigO\left(k^2\right)$ arithmetic operations.
\end{proposition}
The \textsc{Unranking} algorithm computes a run given its rank in the
following total order:
{\small
\[
  \alpha \preceq_{n,k} \beta \quad \text{iff.} \quad 
  \left\{
    \begin{array}{lclc}
      \alpha \in I_{n,k,i_0} \text{ and } \beta \in I_{n,k,i_1} & \land &  i_0 < i_1, & \text{or}\\[5pt]
      \alpha,\beta \in I_{n,k,i} & \land & \begin{array}{l} \alpha
        \text{ is built recursively from } I_{n,k-1,i-2} \text{ and } \\
      	\beta \text{ is built recursively from } I_{n+1,k-1,i-1} \end{array} & \text{or}\\[5pt]
      \alpha,\beta \in I_{n,k,i} & \land & \begin{array}{l} \alpha, \beta  \in I_{n,k-1,i-2} \text{ (resp. } I_{n+1,k-1,i-1} \text{) and } \\
      \alpha_0, \beta_0 \text{ inducing } \alpha,\beta \text{ satisfy } \alpha_0 \preceq_{n,k-1} \beta_0.  \end{array} &
    \end{array}
  \right.
\]
}
\begin{algorithm}[h]
	\caption{\label{algo:unrank} Unranking for $I_{n,k}$}
	\begin{algorithmic}[1]
		\Function{Unranking}{$n,k,r$}
		\State $\ell := k+1$
		\While{$r\geq 0$}
		\State $r := r-t_{n,k,l}$
		\State $\ell := \ell +1$
		\EndWhile
		\State \textbf{return} $\textsc{Cons}(n,k,\ell,r)$
		\EndFunction		
		
		\Function{Cons}{$n,k,\ell,r$}
		\If{$k=0$}
		\State \textbf{return} $\langle x_1, x_2, \dots, x_n \rangle$
		\EndIf
		
		\If{$r < (\ell-2)\cdot t_{n,k-1,\ell-2}$} \Comment generation in $I_{n,k-1,\ell-2}$
		\State $rr := r ~\%~ t_{n,k-1,\ell-2}$
		\State $U := \textsc{Cons} (n, k-1, \ell-2, rr)$
		\State $p_{b} := 1 + r // t_{n,k-1, \ell-2}$ \Comment The position of the new $b$ to insert
		
		\State In $U$, rename each $a_i$ (resp. $c_i$ and $b_i$) by $a_{i+1}$ (resp. $c_{i+1}$ and $b_{i+1}$)
		\State Rename $x_{n-k+1}$ by $c_1$
		\State Insert $b_1$ at position $p_b$ ; and $a_1$ at the head of $U$
		\Else \Comment generation in $I_{n+1,k-1,\ell-1}$
		
		\State $r' := r - (\ell-2)\cdot t_{n,k-1,\ell-2}$
		\State $rr := r' ~\%~ t_{n+1,k-1,\ell-1}$
		\State $U := \textsc{Cons}(n+1, k-1, \ell-1, rr)$
		\State $p_{b} := 2 + r' // t_{n+1,k-1, \ell-1}$
				
		\State Rename $x_{p_b}$ by $b_1$ and $x_{n-k+2}$ by $c_1$ ; and each $x_i$ with $i > p_b$ by $x_{i-1}$
		\State Insert $a_1$ at the head of $U$
		\EndIf
		\State \textbf{return} $U$
		\EndFunction
	\end{algorithmic}
	{\footnotesize
	Line $11$ and $19$ : the binary operator $\%$ denotes the Euclidean division remainder.
	}
\end{algorithm}

The run example of Fig.~\ref{fig:run} has rank $479$ among the
$1270$ runs of $A_{5,4}$.
Note that in the case $k=n$ (at the end there is no $x_1$) the algorithm is easily
extended by considering the position of $b_1$ as the one of $x_1$.
\begin{theorem}
	The Algorithm \textsc{Unranking}$(n,k,r)$ builds the $r$-th run of $A_{n,k}$
	in $k$ recursive calls, once the coefficient memorizations $t_{n,k,\ell}$, for all $\ell$ such that $k+1 \leq \ell \leq 2k+1$
	(and the necessary $n$ and $k$), have been done.
\end{theorem}
Note that the implementation of both algorithms can be much more efficient than the pseudocode exhibited above.
Actually, only the absolute positions of the $b_i$'s are important in a run, because all other actions have
their positions determined by $b_i$'s positions. However, such
implementations are much more cryptic to read, and so we preferred to
present here easy-to-read algorithms.

	\newpage

\begin{thebibliography}{BDGP17b}
	
	\bibitem[AS64]{AS64}
	M.~Abramowitz and I.~A. Stegun.
	\newblock {\em Handbook of Mathematical Functions with Formulas, Graphs, and
		Mathematical Tables}.
	\newblock Dover, New York, ninth dover printing, tenth gpo printing edition,
	1964.
	
	\bibitem[BDF{\etalchar{+}}16]{diamonds16}
	O.~Bodini, M.~Dien, X.~Fontaine, A.~Genitrini, and H.-K. Hwang.
	\newblock Increasing diamonds.
	\newblock In {\em Latin American Symposium on Theoretical Informatics}, pages
	207--219. Springer, Berlin, Heidelberg, 2016.
	
	\bibitem[BDGP17a]{BDGP17bis}
	O.~Bodini, M.~Dien, A.~Genitrini, and F.~Peschanski.
	\newblock Entropic uniform sampling of linear extensions in series-parallel
	posets.
	\newblock In {\em 12th International Computer Science Symposium in Russia
		(CSR)}, pages 71--84, 2017.
	
	\bibitem[BDGP17b]{BDGP17}
	O.~Bodini, M.~Dien, A.~Genitrini, and F.~Peschanski.
	\newblock {The Ordered and Colored Products in Analytic Combinatorics:
		Application to the Quantitative Study of Synchronizations in Concurrent
		Processes}.
	\newblock In {\em 14th SIAM Meeting on Analytic Algorithmics and Combinatorics
		(ANALCO)}, pages 16--30, 2017.
	
	\bibitem[BGP13]{BGP13}
	O.~Bodini, A.~Genitrini, and F.~Peschanski.
	\newblock The combinatorics of non-determinism.
	\newblock In {\em FSTTCS'13}, volume~24 of {\em LIPIcs}, pages 425--436.
	Schloss Dagstuhl, 2013.
	
	\bibitem[BGP16]{BGP16}
	O.~Bodini, A.~Genitrini, and F.~Peschanski.
	\newblock {A Quantitative Study of Pure Parallel Processes}.
	\newblock {\em Electronic Journal of Combinatorics}, 23(1):P1.11, 39 pages,
	(electronic), 2016.
	
	\bibitem[FS09]{FS09}
	P.~Flajolet and R.~Sedgewick.
	\newblock {\em Analytic Combinatorics}.
	\newblock Cambridge University Press, 2009.
	
	\bibitem[FZVC94]{FZVC94}
	P.~Flajolet, P.~Zimmermann, and B.~Van~Cutsem.
	\newblock A calculus for the random generation of labelled combinatorial
	structures.
	\newblock {\em Theoretical Computer Science}, 132(1-2):1--35, 1994.
	
	\bibitem[GS05]{GrSm04}
	R.~Grosu and S.~A. Smolka.
	\newblock Monte carlo model checking.
	\newblock In {\em TACAS'05}, volume 3440 of {\em LNCS}, pages 271--286.
	Springer, 2005.
	
	\bibitem[Knu98]{Knuth98}
	D.~E. Knuth.
	\newblock {\em The art of computer programming, volume 3: (2nd ed.) sorting and
		searching}.
	\newblock Addison Wesley Longman Publishing Co., Inc., Redwood City, CA, USA,
	1998.
	
	\bibitem[KP15]{KP15}
	A.~Khoroshkin and D.~Piontkovski.
	\newblock On generating series of finitely presented operads.
	\newblock {\em Journal of Algebra}, 426:377 -- 429, 2015.
	
	\bibitem[Lip88]{Lipshitz88}
	L.~Lipshitz.
	\newblock The diagonal of a d-finite power series is d-finite.
	\newblock {\em Journal of Algebra}, 113(2):373 -- 378, 1988.
	
	\bibitem[MM03]{MM03}
	C.~Mart{\'i}nez and X.~Molinero.
	\newblock Generic algorithms for the generation of combinatorial objects.
	\newblock In {\em MFCS'03}, pages 572--581. Springer Berlin Heidelberg, 2003.
	
	\bibitem[Nic06]{Nickalls06}
	R.W.D. Nickalls.
	\newblock Viète, descartes and the cubic equation.
	\newblock {\em The Mathematical Gazette}, 90(518):203–208, 2006.
	
	\bibitem[NW75]{NW75}
	A.~Nijenhuis and H.S. Wilf.
	\newblock {\em Combinatorial algorithms}.
	\newblock Computer science and applied mathematics. Academic Press, New York,
	NY, 1975.
	
	\bibitem[Sta01]{Stanley01}
	R.P. Stanley.
	\newblock {\em Enumerative Combinatorics:}.
	\newblock Cambridge Studies in Advanced Mathematics. Cambridge University
	Press, 2001.
	
\end{thebibliography}
\newcommand{\etalchar}[1]{$^{#1}$}

\end{document}